\documentstyle[12pt]{article}
\def\hb{\hbar}  % 
\def\al{\alpha}

\def\rh{\rho}

\def\ph{\phi}

\def\ps{\psi}
\def\om{\omega}

\def\De{\Delta}

\def\Up{\Upsilon}

\def\Ps{\Psi}
\def\Om{\Omega}

\def\fr#1#2{{{#1} \over {#2}}}
\def\prt{\partial}
\def\ap{\al^\prime}

\def\ket#1{|{#1}\rangle}

\def\abs#1{\left|{#1}\right|}                    %      
\def\vect#1{{\bf #1}}  % \vec will make an arrow over the argument

\def\half{{\textstyle{1\over 2}}}

\def\quar{{\textstyle{1\over 4}}}  % 
\def\frac#1#2{{\textstyle{{#1}\over {#2}}}}

\def\lsim{\mathrel{\rlap{\lower4pt\hbox{\hskip1pt$\sim$}}
    \raise1pt\hbox{$<$}}}
\def\gsim{\mathrel{\rlap{\lower4pt\hbox{\hskip1pt$\sim$}}
    \raise1pt\hbox{$>$}}}
\def\sqr#1#2{{\vcenter{\vbox{\hrule height.#2pt
         \hbox{\vrule width.#2pt height#1pt \kern#1pt
         \vrule width.#2pt}
         \hrule height.#2pt}}}}

\def\x{{\bf \hat e}_x} %
\def\y{{\bf \hat e}_y} %
\def\z{{\bf \hat e}_z}
\def\phihat{{\bf \hat e}_\phi} %
\def\rhohat{{\bf \hat e}_\rho} %
\def\Lgr#1#2#3{L_{\scriptstyle #1}^{(\scriptstyle #2)}\left(#3\right)}

\newcommand{\beq}{\begin{equation}}
\newcommand{\eeq}{\end{equation}}
\newcommand{\bea}{\begin{eqnarray}}
\newcommand{\eea}{\end{eqnarray}}
\newcommand{\rf}[1]{(\ref{#1})}
\newcommand{\eq}[1]{Eq.\ \rf{#1}} %
\renewenvironment{thebibliography}[1]
 { \rm
   \begin{list}{\arabic{enumi}.}
    {\usecounter{enumi} \setlength{\parsep}{0pt}
     \setlength{\itemsep}{3pt} \settowidth{\labelwidth}{#1.}
     \sloppy
    }}{\end{list}}

\begin{document}

\titlepage

\begin{flushright}
{IUHET 352\\}
{December 1996\\}
\end{flushright}
\vglue 1cm
	    
\begin{center}
{
{\bf ANALYTICAL MODELS FOR VALENCE FERMIONS\\ }
{\bf IN ISOTROPIC TRAPS\\ }
\vglue 1.0cm
{V. Alan Kosteleck\'y and Neil Russell\\} 
\bigskip
{\it Physics Department, Indiana University,\\}
\medskip
{\it Bloomington, IN 47405, U.S.A.\\}

\vglue 0.8cm
}
\vglue 0.3cm

\end{center}

{\rightskip=3pc\leftskip=3pc\noindent

For isotropic confining Ioffe-Pritchard or TOP potentials,
a valence fermion trapped with a closed core of other fermions
can be described by an analytical effective one-particle model
with a physical eigenspectrum.
Related constructions exist for Paul and Penning traps.
The analytical models arise from quantum-mechanical supersymmetry.

}


\vskip 1truein
\centerline{\it Accepted for publication 
in Physics Letters A}

\vfill
\newpage

\baselineskip=20pt

Electromagnetic traps may be divided into two categories:
those that trap neutral particles or atoms, 
and those that trap charged particles or ions.
The trapping mechanism for the former
usually exploits the force experienced by a magnetic dipole
in a nonuniform magnetic field.
Typical configurations consist of a magnetic quadrupole field
supplemented with a mechanism to reduce trapping losses 
near the field zero.
Two well-established examples are the Ioffe-Pritchard trap 
\cite{ioff,prit},
which has end coils ensuring a nonzero-field minimum,
and the time-averaged orbiting-potential (TOP) trap 
\cite{paec},
which employs a rotating magnetic field.
Discussions of various neutral-particle magnetostatic traps 
are given in Ref.\ \cite{bem}.
Charged particles and ions can be trapped 
with the Paul trap 
\cite{wp},
which involves an oscillating electric potential
and can simultaneously trap particles of both polarities.
An alternative is the Penning trap 
\cite{fmp,deh1},
which combines electrostatic and magnetic fields
and exists in many forms 
\cite{bg,grhk,pa}.
A general discussion of ion traps is given in Ref.\ \cite{pkg}.

The quantum behavior of many particles confined in a trap
has been the subject of much recent experimental and 
theoretical work
\cite{aemwc,kb}.
Most investigations have centered on
systems involving large numbers of bosons.
In the present paper,
we address instead some quantum-physics issues
for certain traps containing 
one to a few hundred \it fermions, \rm
with the system in the ground state.
We seek to describe the physics 
of an additional valence fermion added to such a system,
perhaps in a highly excited state,
using a relatively simple analytical model.
Note that the focus of the present work is 
primarily theoretical issues involving quantum physics.
In particular, 
we disregard experimentally important issues 
such as thermal effects
\cite{hw,kb}.

Part of our interest in situations of this type 
stems from the analogy with Rydberg atoms,
which have played a central role 
in the development of tools 
for understanding multi-electron atoms
\cite{gallagher}.
Certain related developments could emerge in the context
of multiparticle traps of the type we consider.
Rydberg states of alkali-metal atoms are widely used
for experimental and theoretical investigations,
in part because 
the associated electronic core is relatively simple,
forming a closed shell.
For related reasons,
in the present work we primarily consider systems
of trapped fermions with the valence fermion
lying outside a closed fermionic shell.
The analogy with alkali-metal atoms is best 
for traps with a confining potential
that is purely isotropic in three dimensions.
For certain values of the applied fields,
the Ioffe-Pritchard and TOP traps provide examples
that closely approximate this situation. 

Both the Ioffe-Pritchard and TOP traps 
capture magnetic dipoles $\vec \mu$
aligned oppositely to an applied magnetic field $\vect B$
by drawing them into a region of minimum $\abs {\vect B}$.
The potential energy is $U(x,y,z) = \mu\abs{\vect B}$.
Dipoles aligned with the field are expelled from the trap.

The Ioffe-Pritchard trap consists of two coils 
with aligned symmetry axes in,
say, the $\z$ direction
and four conducting bars equidistant from 
and parallel to the $z$ axis. 
The coils carry parallel currents,
while the bars carry alternately oriented currents.
In cylindrical coordinates,
the magnetic field to third order in $\rh$ and $z$ is
\beq
{\vect B}_{\mbox{\tiny IP}} (\rho,\phi,z) = B_c \z 
+B_l^\prime\rho\left[\cos{2\phi}\rhohat-\sin{2\phi}\phihat\right]
+\fr{B_c a_c}{l_c^4}\left[(z^2-\rho^2/2)\z-z\rho\rhohat\right]  
\label{ip1} 
\quad .
\eeq
In this expression,
the coils are taken to have radius $R_c$,
to carry currents $I_c \phihat$,
and to be positioned at $z= \pm  A_c$.
The bars are assumed to be located at 
$\rh = S_l$ with $\phi = \pi/4$, 
$3\pi/4$, $5\pi/4$, $7\pi/4$,
and to carry currents of magnitude $I_l$. 
The other quantities are a characteristic length,
area, field, and field gradient:
$l_c=(A_c^2+R_c^2)^{1/2}$,
$a_c = 3(4A_c^2-R_c^2)/2$,  
$B_c=\mu_0 I_c R_c^2/l_c^3$,
$B_l^\prime=2\mu_0 I_l /\pi S_l^2$.
Trapped magnetic dipoles have potential energy 
\beq
U_{\mbox{\tiny IP}}(\rho,\phi,z) 
= \mu B_c \left[1+ \fr{a_c}{l_c^4}z^2
+ \fr 1 2 \left(\fr{{B_l^\prime}^2}{B_c^2}
        -\fr{a_c}{l_c^4}\right)\rho^2\right]
\label{ip2}
\quad .
\eeq
If the condition 
${B_l^\prime}^2/B_c^2= 3 a_c/l_c^4 $
is satisfied,
as is readily feasible in practical situations, 
then the potential \eq{ip2} is isotropic:
\beq
U_{\mbox{\tiny IP}}(r) 
= \mu B_c \left( 1+ \fr{r^2}{r_c^2} \right)
\label{ip4}
\quad ,
\eeq
where $r_c^2=l_c^4/a_c$.
Note that the orientation of the dipoles is anisotropic,
for example,
lying roughly along the $-\z$ direction near the $z$ axis.

The TOP trap 
involves two parallel coils as above,
but carrying oppositely oriented currents,
and two further pairs of coils
providing an additional rotating bias field.
Let the quadrupole coils 
have radius $R_q$,
be located at $z= \pm A_q$,
and carry currents $\mp I_q \phihat$,
respectively.
Let the pairs of bias coils 
of radius $R_b$
be located at $x=\pm A_b$ on the $x$-axis
and at $y=\pm A_b$ on the $y$-axis,
and let the corresponding currents be
$I_x(t) = I_b \cos{\omega_b t}$
and $I_y(t) = I_b \sin{\omega_b t}$.
The time dependence 
creates a magnetic field vector
that to lowest order lies parallel to the $z=0$ plane
and rotates with frequency $\omega_b$.
To second order in $x$, $y$, and $z$,
the magnetic field is
\bea
{\vect B}_{\mbox{\tiny TOP}}(x,y,z,t) &=&    
 \left( B_b \cos{w_b t} 
  + B_q^\prime x 
  + \fr{B_b a_b}{2 l_b^4} 
     \left [(3x^2-r^2)\cos{\omega_b t} 
     - 2xy \sin{\omega_b t} \right ]
 \right) \x \nonumber\\
&+& \left( B_b \sin{\omega_b t} 
  + B_q^\prime y 
  + \fr{B_b a_b}{2 l_b^4} 
     \left [(3y^2-r^2)\sin{\omega_b t} 
     - 2xy \cos{\omega_b t} \right ]
 \right) \y \nonumber\\
&-& \left(  2 B_q^\prime z 
         + \fr {B_b a_b}{l_b^4}
            \left [ x \cos{\omega_b t} 
	    + y \sin{\omega_b t}\right ] z
  \right) \z
\quad ,
\label{top1} 
\eea
where the characteristic length, area, field, 
and field gradient are
$l_b = (A_b^2 + R_b^2)^{1/2}$,
$a_b= 3(4A_b^2-R_b^2)/2$, 
$B_b= \mu_0 I_b R_b^2 / l_b^3$,
$B_q^\prime = 3\mu_0 I_q R_q^2 A_q / 2 (A_q^2+R_q^2)^{5/2}$. 
The potential energy is obtained by expanding 
$\mu \abs{\vect B_{\mbox{\tiny TOP}}}$ 
for small $\rh$, $z$ and time averaging,
assuming that the frequency is high enough so the dipoles 
do not move appreciably over one cycle.
The time averaging removes linear and cubic terms,
producing the trapping potential
\beq
U_{\mbox{\tiny TOP}}(\rho, z) =
 \mu B_b\left[1
   +\fr 1 4 \left(\fr{{B_q^\prime}^2}{B_b^2}
   +\fr{a_b}{l_b^4}\right)\rho^2 
   +\left(\fr{2{B_q^\prime}^2}{B_b^2}-\fr{a_b}{2l_b^4}\right)z^2
  \right] 
\label{top2}
\quad 
\eeq
to third order.
If the physical parameters satisfy
${B_q^\prime}^2/B_b^2= 3a_b/7l_b^4 $,
a readily attainable condition,
then the potential becomes isotropic:
\beq
U_{\mbox{\tiny TOP}}(r) =
 \mu B_b\left(1 + \fr{r^2}{r_b^2} \right)
\label{top4}
\quad ,
\eeq
where $r_b^2 = 14l_b^4/5a_b$.
Note that the dipole orientation is anisotropic,
being opposite to the time-averaged field \eq{top1} and
pointing away from the origin along the $z$ axis 
and radially towards the origin in the $z=0$ plane.

The Schr\"odinger equation for 
a particle in either isotropic trap is
\beq
\left(-\fr{\hbar^2}{2m} \vect\nabla^2 +\mu B_0
+ \mu B_0 \fr{r^2} {r_s^2} \right) \psi(r,\theta,\phi) 
 = E \psi(r,\theta,\phi) 
\quad ,
\label{ho1}
\eeq
where $m$ is the dipole mass, 
$B_0 = B_c$ or $B_b$,
and $r_s = r_c$ or $r_b$.
Separating in spherical polar coordinates,
we write
$\psi(r,\theta,\phi)=(r_0/r)W(r)Y(\theta,\phi)$,
where $Y(\theta,\phi)$ are spherical harmonics,
$r_0=(\hbar/m \om_0)^{1/2}$,
and $\om_0=(2\mu B_0/m r_s^2)^{1/2}$.
This choice puts the radial equation 
for $W(r)$ into a convenient form:
\beq
\left(-\fr{\hbar^2}{2m} \fr{d^2}{dr^2}
     + \fr{\hbar^2}{2m} \fr{L(L+1)}{r^2}
     + \mu B_0+\half m \om_0^2 r^2 \right) W(r)
 = E_N W(r)
\quad ,
\label{ho2}
\eeq
where the angular momentum quantum number $L$
takes values $L=0, 1, 2, \ldots$
and the energy eigenvalues are
$E_N = \mu B_0+\hbar\om_0(N+3/2)$,
with principal quantum number $N=L,L+2,L+3,\ldots$.
The wave functions are
given in terms of generalized Laguerre polynomials:
\beq
W_{N,L}(r) = C_{N,L} (r/r_0)^{L+1} \exp(-r^2/2 r_0^2)
     \Lgr{N/2 - L/2}{L + 1/2}{r^2/r_0^2}
\quad ,
\label{ho4}
\eeq
where the quantities $C_{N,L}$ are normalization constants.

In what follows,
it is useful to consider
situations where a cloud of fermions 
forms a closed shell of energy levels in an isotropic trap.
The number of fermions forming a closed shell 
below a particular value of $N$
is determined by the degeneracies of states 
in the single-particle system.
Examining the full wave functions 
and recalling that the allowed values of $N$
increase in steps of two units,
the degeneracy for a given $N$
is found to be $(N+1)(N+2)/2$.
Note that the doubling in atomic systems 
due to the two spin orientations 
has no analogue here because only dipoles 
oriented against the magnetic field are trapped.
The number of particles completely filling levels
less than or equal to $N$ is
$(N+1)(N+2)(N+3)/6$.

In the present work,
we seek to construct relatively simple 
analytical one-particle models for the valence fermion
in cases with more than one trapped particle.
In addition to their intrinsic interest,
such models could be used to make 
analytical predictions of physical properties
or could provide a favorable starting point
for perturbative and other calculations.
The methodology applied here to generate effective models
for various trap systems 
is related to that adopted in the development of
a relatively simple analytical model for 
the valence electron in Rydberg atoms 
\cite{kn}.
This model has been used in a variety of contexts
\cite{ak,bk},
including recently the prediction 
of certain experimentally observable features of 
long-term revivals in Rydberg atoms.

Consider a system of fermions caught in a trap,
with one particle excited relative to the others.
This valence fermion can be regarded as moving
in an effective potential 
created by a combination of the trapping fields
and interactions with the other trapped particles.
We include among these interactions 
the quantum effects from the Pauli principle,
which prevents the valence fermion from 
occupying filled levels,
and also interparticle forces 
that act to modulate the trapping fields.
In constructing analytical models for the valence fermion,
we address first the issue of incorporating effects 
from the Pauli principle 
and subsequently examine an analytical extension
that could describe other interparticle forces.

To illustrate the idea,
consider the particular tower of states 
$\ket{N,L=0,M=0}$
accessible to a trapped valence fermion with 
angular quantum number $L=0$
and azimuthal quantum number $M=0$.
If only one particle is trapped,
then $N\ge 0$ with $N$ even.
If there is also a core of four fermions 
filling the energy levels below $N=2$,
then a valence fermion with $L=0$
is restricted by the Pauli principle
to levels with even $N\ge 2$. 
Neglecting for the moment other interactions
among the trapped particles,
the valence fermion in each of the two situations
can access states with identical eigenenergies,
except that in the five-particle case 
the $N=0$ level is inaccessible. 
The issue of constructing 
an analytical effective one-particle theory
to describe the $L=0$ states of the  
valence fermion in the five-particle case
can therefore be rephrased as the problem of finding 
an analytically solvable effective trapping potential
with $L=0$ energy eigenstates identical 
to those of the one-particle case
but with the lowest one-particle level missing.

Isospectral problems of this type can be treated in several ways.
One approach might be the inverse method 
\cite{glam}.
In the present work,
we choose instead an alternative method 
with a definite physical interpretation
that is both elegant and relatively simple,
based on supersymmetric quantum mechanics
with the superalgebra sqm$(2)$
\cite{kn,ak}.
This supersymmetric technique takes as input 
a Schr\"odinger hamiltonian $H^+$ 
for which the ground-state eigenenergy vanishes,
and it determines via supersymmetry a complementary hamiltonian $H^-$.
By construction,
the eigenstates of $H^-$ are
degenerate with those of the original potential
except for the ground state, which is absent.
The eigenstates of the two systems $H^+$ and $H^-$ 
are related by an explicit map.
The two associated Schr\"odinger equations
can be written as
\beq
H^\pm \ps^\pm \equiv \fr{\hbar^2}{2m} 
\left( 
- \fr{d^2}{dr^2} 
+ \left(\fr{dU}{dr}\right)^2 \mp \fr{d^2 U}{dr^2}
\right) \ps^\pm
= E^\pm \ps^\pm
\quad ,
\label{susy4}
\eeq
where $\ps^\pm$ are the eigenstates of $H^\pm$
with eigenvalues $E^\pm$.
The combinations of derivatives of $U(r)$
generate associated potentials $V^\pm (r)$.
Since by assumption $H^+$ and hence $V^+(r)$ are given,
the function $U(r)$ can be found by 
solving a differential equation.
The form of $V^-(r)$ and $H^-$ can then be deduced.
The existence of the degeneracy and explicit map 
between excited states in the two eigenspectra
is a direct consequence of the supersymmetry.

This procedure can be applied to Eq.\ \rf{ho2}.
Subtracting the ground-state energy 
$E_L = \mu B_0+\hbar\om_0(L+3/2)$
makes the lowest eigenvalue of the radial hamiltonian vanish, 
as required by the method.
This subtraction depends on the angular momentum $L$,
which therefore must be fixed.
The input potential becomes
\beq
V_L^+(r) = \fr{\hbar^2}{2m} \fr{L(L+1)}{r^2}
     +\half m \om_0^2 r^2 
     - \fr{\hbar\om_0}{2}(2L+3)
\quad ,
\label{qs1}
\eeq
with eigenenergies
$E_{N,L}^+ = \hbar\om_0(N-L)$
and eigenfunctions $W_{N,L}^+(r) =W_{N,L}(r)$.
The solution for the function $U(r)$ is
\beq
U(r) = \fr 1 2 \left(\fr r {r_0}\right)^2 
      - (L+1) \ln\left(\fr r {r_0}\right)
\quad ,
\label{qs3}
\eeq
from which we obtain the form of $H^-$:
\beq
\left( - \fr{\hbar^2}{2m} \fr{d^2}{dr^2}
       + \fr{\hbar^2}{2m} \fr{(L+1)(L+2)}{r^2}
       + \half m \om_0^2 r^2 
       - \fr{\hbar\om_0}{2} (2L+1)
      \right) W_{N_s, L}^-
 = E_{N_s, L}^- W_{N_s, L}^-
\quad .
\label{qs4}
\eeq 
The energy eigenvalues are
$E_{N_s,L}^- = \hbar\om_0(N_s-L)$,
where $N_s = L+2, L+4, \ldots$,
and the eigenfunctions are 
$W_{N_s, L}^-(r) = W_{N_s-1, L+1}(r)$.

The derivation shows that
for the $L=0$ case discussed above
the eigenfunctions 
$(r_0/r) W_{N_s, L=0}^-(r) Y_{L=0,M=0}(\theta,\phi)\equiv
(r_0/r) W_{N_s-1, L=1}(r) Y_{L=0,M=0}(\theta,\phi)$
are effective one-particle eigenfunctions 
for the valence particle 
in a trap containing a total of five fermions.
We emphasize that these 
are \it not \rm conventional oscillator eigenfunctions,
as can be seen from the index structure.

It might be tempting instead to model the valence fermion 
directly using conventional oscillator wave functions,
based on a shell-model approach where the Pauli 
principle is incorporated by hand.
However,
the supersymmetric eigenfunctions have 
several advantages.
Unlike the conventional oscillator case,
for which the lowest state must be excluded by hand, 
the supersymmetric states form a complete set.
Moreover,
the lowest-state radial eigenfunction
$W_{N_s=2, L=0}^-(r)$
in our effective model 
has degree zero and hence zero nodes,
as expected for the ground state of the five-fermion system.
A conventional oscillator wave function would have one node instead.

The potentials $V_0^-$ and  $V_0^+$ differ 
by an inverse-square repulsive term:
\beq
V_0^-(r) - V_0^+(r)
= \fr{\hbar^2}{m} \fr 1 {r^2} + \hbar\om_0 
\label{qs7}
\quad .
\eeq
In the present context,
the additional repulsion in $V_0^-$
plays the role of the Pauli principle
by preventing the valence fermion
from occupying the filled lower levels.
Note that a change in the angular-momentum barrier
would produce a similar effect on the potential
but would not connect states with the same value of $L$.

The above arguments for the case $L=0$
can also be applied to other values of $L$.
For example,
in the case where the valence fermion has $L=1$
the method produces a relation between
two effective one-particle models,
the first involving two trapped fermions 
with one in the ground state
and the second involving eleven trapped fermions 
with ten filling the levels below $N=3$.

Iterations of the method produce further relations.
When the valence fermion has $L=0$,
for instance,
the effective one-particle model for the
system with five trapped fermions
described by Eq.\ \rf{qs4}
can in turn be related to another model
for a system with 21 trapped fermions,
20 of which fill the levels below $N=4$. 
Implementing this mathematically
requires another shift of the energy zero
so that the ground state of the five-particle system
has zero energy.
Repeating the procedure produces a series of
interrelations between effective one-particle models
for the $L=0$ towers of states 
of systems with $n= 1, 5, 21, 57, \ldots$ trapped fermions.
A similar iteration for $L=1$ generates
connections between systems with 
$n= 2, 11, 36, 85, \ldots$ trapped fermions.
The two sequences of numbers are generated by the
formula $n= 1 + (N+1)(N+2)(N+3)/6$, 
with $N= -1, 1, 3, \ldots$ or $N= 0, 2, 4, \ldots$.

The preceding discussion has largely disregarded 
effects of interactions between the trapped particles.
If particle interactions are entirely neglected,
the degeneracy of each fixed-$N$ level means that 
the same effective one-particle model 
applies to the valence fermion 
in traps containing $n_d = d + (N+1)(N+2)(N+3)/6$
particles,
where $d = 1, 2, \ldots, (N+1)(N+2)/2$
is the number of fermions lying outside a closed shell. 
Although there may be special situations where 
particle interactions are relatively small,
the possibility of additional interactions among the $d$
valence fermions suggests that the best effective models
would typically have $d=1$.
Similarly,
the best effective models should also be ones describing 
small numbers of trapped particles 
and a relatively highly excited valence fermion.
These considerations favor, 
for example, 
the model for the five-particle case
with one highly excited fermion.

Issues involving particle interactions
could be addressed using a variety of standard methods,
such as mean-field theory, perturbation theory,
and Monte-Carlo methods.
Given our present focus,
we pursue here instead the possibility 
of incorporating interactions via analytical modifications 
to the effective models.

Interactions typically shift the energy eigenvalues
of the valence fermion.
We treat this as a shift $\De=\De(N,L)$ 
in the principal quantum number.
Since the effective model produces 
apparent integer shifts in $L$,
we define for convenience an integer $I=I(L)$ 
and introduce the effective principal quantum number
$N^* = N+I-\De$.
It is also convenient to define $N_s=N+2I$,
a quantum number analogous 
to the spectroscopic principal quantum number 
in atomic systems.
We therefore have $N^* = N_s-I-\De$,
with new energy eigenvalues
$E_{N^*} = \mu B_0 + \hb\om_0\left(N^* + 3/2 \right)$. 

The problem is to find a modification 
of the radial potential
in the above effective one-particle models
such that the eigensolutions 
of the corresponding radial equation
remain analytical but are associated with 
the modified eigenenergies $E_{N^*}$.
This minor miracle can be accomplished by
adding the effective potential
\beq
V_{\mbox{\tiny EFF}}(r) 
= \fr{\hb^2}{2m} \fr{L^*(L^* +1)-L(L+1)}{r^2}
+ \hb\om_0(N-N^*)
\quad 
\label{qdt1}
\eeq
to the operator on the left-hand side of \eq{ho2},
where $L^*=L+I-\De$.
The resulting differential equation,
\beq
\left( -\fr{\hbar^2}{2m} \fr{d^2}{dr^2}
     + \fr{\hbar^2}{2m} \fr{L^*(L^*+1)}{r^2}
     + \mu B_0+\half m \om_0^2 r^2 \right) W(r)
 = E_{N^*} W(r)
\quad ,
\label{qdt2}
\eeq
has eigenvalues $E_{N^*}$ with
$N^*=L^*$, $L^*+2$, $L^*+4$, $\ldots$.
In terms of the functional form given in Eq.\ \rf{ho4},
the eigensolutions are $W_{N^*,L^*}(r)$.
Note that 
the special case $\De=0$ reproduces all
the complementary hamiltonians $H^\pm$ discussed above
when $\mu B_0 + \hb\om_0(L^* + 3/2 -2S)$
is subtracted from both sides of Eq.\ \rf{qdt2}.
Thus,
the radial equation for $H^+$ is recovered
by selecting $S=0$ and $I=0$,
while that for $H^-$ is recovered
by selecting $S=1$ and $I=1$.
The $I_0$th iteration of $H^+$ is obtained with
$S=0$ and $I=I_0$,
while the corresponding $H^-$ has
$S=1$ and $I=I_0+1$.

The analysis given here has some similarities to 
the derivation of a relatively simple analytical model 
for the valence electron in Rydberg atoms discussed in 
Refs.\ \cite{kn,ak,bk}.
For example,
our treatment of the $L=0$ states of isotropic trap systems 
with 1, 5, 21, 57, $\ldots$ fermions
resembles that of the s-orbital states of alkali-metal atoms
in the Rydberg case.
Note that,
since in the present trap context the valence fermion
is typically neutral,
its interactions involve a dipole moment
rather than a monopole charge.
This suggests that the effect of interactions
may be significantly smaller than in the Rydberg case 
and the effective models correspondingly better.
Note also that the form of the effective potential \rf{qdt1}
is reminiscent of the special analytical model investigated
in Ref.\ \cite{bqj}.

In the atomic case,
orthogonality of the model eigenfunctions arises 
because the quantum defects in, 
for example,
alkali-metal atoms are asymptotically independent 
of the principal quantum number.
Similarly, 
in the present context
the eigenfunctions $W_{N^*,L^*}(r)$
form an orthonormalizable basis 
provided $\De$ is independent of $N$ 
with $\De < L + I + 3/2$.
The $N$-independence implies that for each fixed $L$ 
only one parameter is needed, 
corresponding to a simultaneous shift 
of the principal quantum numbers for that tower of states.
Experience gained in the atomic case suggests that
rapid asymptotic $N$-independence
is likely to suffice for applicability of the models here.
However,
it is unclear \it a priori \rm
which trap systems have this feature,
as it depends on details of the many-body dynamics.
Experimental investigations establishing
energy spectra for the systems considered here 
would be of interest.
The eigenfunctions $W_{N^*,L^*}(r)$ could then be used 
to predict other quantities such as transition rates. 

In contrast to the Ioffe-Pritchard or TOP traps,
which can be isotropic in three dimensions,
the Paul and Penning traps are generically
isotropic in two dimensions.%
\footnote{
In a particular rotating frame,
a single particle in a Penning trap
with a special ratio of applied fields
can experience an isotropic potential in three dimensions.
The incorporation of rotation effects,
including those on any fermion core,
could then allow a treatment similar 
to that for the Ioffe-Pritchard and TOP traps.}
Effective one-particle models can nonetheless
be obtained via a similar approach.
For brevity,
we restrict ourselves to outlining
the treatment of the case where interactions
of the valence fermion are neglected.
Effective one-particle models 
analogous to Eq.\ \rf{qdt2}
that allow for level shifts
can also be constructed for the Paul and Penning traps.

The Paul trap consists of a time-dependent potential
given in cylindrical coordinates by
$\tilde{\ph}(\rh,\ph,z,t) = \ph(\rh,\ph,z) \, 
\cos{\tilde\Om_p t}$,
where $\ph$ is
\beq
\phi(\rh,\ph,z) = \fr{V_p}{2d_p^2}(z^2- \rh^2/2)
\quad 
\label{pl2}
\eeq
with characteristic voltage $V_p$ and length $d_p$.
A quantum solution for the one-particle case exists
\cite{lb}
and can be investigated experimentally 
\cite{bls},
but for simplicity we consider here 
an alternative approach for large 
$\tilde\Om_p \gg \Om_p \equiv (\sqrt{2}~|q V_p|/ m d_p^2)^{1/2}$
in which $q\tilde\ph$ is approximated by a time-independent 
effective potential
\cite{csw}
\beq
\overline{V}
\equiv \fr{q^2}{4 m \tilde\Om_p^2} 
\vect \nabla \ph \cdot \vect \nabla \ph
= \half m \om_p^2 (\rh^2 + 4 z^2)
\quad ,
\label{pl4}
\eeq
where $m$ and $q$ are the mass and charge
of the trapped fermion and $\om_p = \Om_p^2/4\tilde\Om_p$.
The associated quantum problem separates with a wave function
of the form
$\Ps(\rh,\ph,z) = (\rh_p/\rh)^{1/2} X(\rh) \Up(\ph,z)$,
where $\rh_p = (\hb / m \om_p)^{1/2}$.
Introducing the quantum numbers $K=0,1,2,\ldots$ and
$M=0, \pm 1, \pm 2, \ldots$ gives
\beq
\Up_{M,K}(\ph,z) =  
A_K \exp(i M \ph) \exp(- z^2/\rh_p^2) 
H_K(\sqrt 2 ~z/\rh_p)
\quad , 
\label{p19}
\eeq
where the $H_K$ are Hermite polynomials
and the $A_K$ are normalization coefficients.
The two-dimensional radial equation is
\beq
\left( 
- \fr{\hb^2}{2m} \fr{d^2}{d\rh^2} 
+ \fr{\hb^2}{2m} \fr{M^2-\quar}{\rh^2}
+ \half m \om_p^2 \rh^2 
+ \hb \om_p (2K +1) \right) X(\rh) 
= E X(\rh) 
\quad .
\label{pl5}
\eeq
The associated energy eigenvalues are
$E_{N,K} = \hb \om_p (N + 2K + 2)$,
where $N = \abs M, \abs M + 2, \abs M + 4 , \ldots$,
and the eigenfunctions are 
\beq
X_{N,\abs M}(\rh) = 
C_{N,\abs M} \left(\rh /{\rh_p}\right)^{\abs M +1/2}
         \exp (- \rh^2 / 2\rh_p^2)
         \Lgr{N/2-\abs M /2}{\abs M }{\rh^2 /\rh_p^2}
\quad ,
\label{pl7}
\eeq
where the $C_{N,\abs M}$
are normalization coefficients.
 
The Penning trap involves 
an electrostatic field of the form \rf{pl2},
along with a uniform magnetic field $\vect B = B_p \z$.
Defining the axial frequency
$\om_z = (|q V_p|/ m d_p^2)^{1/2}$,
the cyclotron frequency 
$\om_c = \abs{q B_p}/m$,
and $\Om = (\om_c^2-2\om_z^2)^{1/2}$,
the one-particle hamiltonian for $q>0$ is
\beq
H = -\fr{\hb^2}{2m} \vect{\nabla}^2
  + \frac{1}{8} m \Om^2\rh^2 
  + \half m \om_z^2 z^2
  + \half \hbar \om_c i \prt_\ph
\label{pt3}
\quad . 
\eeq
The equation separates via 
$\Psi(\rh,\phi,z)=(\rh_0/\rh)^{1/2}X(\rh)\Theta(\theta,z)$,
where 
$\rh_0=(\hbar/m\om_c)^{1/2}$. 
The $\rh$ equation is
\beq
\left( -\fr{\hbar^2}{2m} \fr{d^2}{d\rh^2}
     + \fr{\hbar^2}{2m}\fr{M^2-\frac 1 4}{\rh^2}
     + \frac{1}{8} m \Om^2\rh^2 
     + (K+\half)\hbar\om_z-\half M\hbar\om_c
\right) X(\rh)
=E X(\rh)
\quad ,
\label{pt4}
\eeq
where $M=0, \pm 1, \pm 2, \ldots$ and $K=0,1,2,\ldots$.
The energy eigenvalues are
$E_{N,K,M} = \hbar (\Om N + 2\om_z K 
- \om_c M +\Om +\om_z)/2$,
where $N= \abs M, \abs M +2,\abs M +4, \ldots $.
The full eigensolutions
involve generalized Laguerre and Hermite polynomials:
\bea
\Psi_{N,K,M}(\rh,\phi,z) &=& C_{N,K,\abs{M}}
    (\rh/\rh_0)^{\abs M}
    \exp{\left[-\frac{k}{4} \left(\frac{\rh}{\rh_0}\right)^2 
        -\half\left(\frac{z}{z_0}\right)^2+iM\phi\right]}
\nonumber \\
&& \qquad \qquad \qquad \qquad
\times\Lgr{N/2-\abs M/2}{\abs M} {k\rh^2/2\rh_0^2} H_K(z/z_0)
\quad ,
\label{pt6}
\eea
where $k=\Om/\om_c$,
$z_0=(\hbar/m\om_z)^{1/2}$,
and $C_{N,K,\abs M}$ are normalization coefficients.

Equations \rf{pl5} and \rf{pt4} are both radial equations 
for a two-dimensional oscillator,
and the same approach to effective one-particle models
applies to each.
For brevity, 
we treat primarily the Paul case in what follows.
Subtracting the ground-state energy, 
the equation analogous to \rf{qs1} is
\beq
V_{\abs M}^+ =   \fr{\hbar^2}{2m} \fr{\abs M ^2-1/4}{\rh^2}
            + \half m \om_p^2 \rh^2
            - \hb\om_p (\abs M +1) 
\quad .
\label{pa1}
\eeq
The partner potential is 
$V_{\abs M}^- = V_{\abs M+1}^+ + 2 \hbar \om_p$.
The difference $V_{\abs M}^- - V_{\abs M}^+ $
again acts as an additional repulsion 
that can be regarded as preventing 
the valence fermion from accessing a filled lower level.
The corresponding partial eigenfunctions are 
$X^+_{N,\abs M}(\rh) = X_{N,\abs M}(\rh)$
and $X^-_{N_s,\abs M}(\rh) = X_{N_s-1,\abs M +1}(\rh)$, 
where $N_s = \abs M +2$, $\abs M +4$, $\ldots$.
The eigenspectra
$E^+_{N,\abs M} = \hb\om_p(N-\abs M)$ and 
$E^-_{N_s,\abs M} = \hb\om_p(N_s-\abs M)$ 
are degenerate,
except for the ground state.
Similar expressions arise for the Penning trap.

For both types of trap,
the potential $V_{\abs M}^+ $ depends on $M$
and the subtracted ground-state energy depends on $K$
(see Eqs.\ \rf{pl5} and \rf{pt4}).
The effective one-particle models 
therefore describe towers of states $\ket{N, M, K}$ 
with fixed $K$ and $M$.
For the Paul trap,
the number of spin-1/2 states
with energy less than or equal to $E_{N,K}$ is
\beq
n(\tilde E) = 
\left\{ 
\begin{array}{ll}
\tilde E(\tilde E +2)(2\tilde E-1)/12 \: , 
& \tilde E \mbox{ even} \: , 
\\
(\tilde E^2 -1)(2\tilde E +3)/12 \: , 
&  \tilde E \mbox{ odd} \: ,
\end{array}
\right.
\label{pa4}
\eeq
where $\tilde E = E_{N,K}/\hb\om_p$.
The Paul-trap systems related
by effective one-particle models
are therefore those with 
$1$, $3$, $7$, $15$, $27$, $45$, 
$69$, $\ldots$ trapped fermions. 

Similar considerations apply for the Penning trap.
Note, 
however,
that in this case the number of spin-1/2 states
with energy less than or equal to $E_{N,M,K}$ 
depends on the frequency tuning.
The magnetron motion is also unstable,
so the corresponding quantum-number combination 
$(N+M)/2$ cannot be too large.

In closing,
we remark that the radial equations 
for all the traps considered in this paper 
can be mapped into various radial equations
for Coulomb-type potentials
\cite{kr},
among which is the usual radial Coulomb equation
in three dimensions.
This suggests that under suitable circumstances
the analogies drawn above 
between atoms and trapped systems
might be more than heuristic.
It would be amusing,
for example,
to identify a trap system 
for which the oscillator energy shifts
correspond via a direct map to 
known quantum defects for an alkali-metal atom.

\vskip 0.4 cm

We thank R. Bluhm, G. Gabrielse, and R. Pollock
for useful comments.
This work was supported in part by the 
United States Department of Energy 
under grant no.\ DE-FG02-91ER40661.
 
% 3 arguments for bib entries : vol, year (yy), page
% for jmp, pra need : J.Name BFvol#, page (year)    (note punctuation)
% for example  : J.Phys. A. 21, 307 (1988)
% refer to bib items with \cite{} 
\def\apb #1 #2 #3 {App. Phys. B {\bf #1}, #3, (19#2)}    %  
\def\ajp #1 #2 #3 {Am.\ J.\ Phys.\ {\bf #1}, #3 (19#2)}
\def\ant #1 #2 #3 {At. Dat. Nucl. Dat. Tables {\bf #1}, #3 (19#2)}
\def\ap #1 #2 #3 {Ann.\ Physics\ {\bf #1}, #3 (19#2)} % 
\def\ijqc #1 #2 #3 {Internat.\ J.\ Quantum\ Chem.\  
  {\bf #1}, #3 (19#2)} %  
\def\jmp #1 #2 #3 {J.\ Math.\ Phys.\ {\bf #1}, #3 (19#2)}
\def\jms #1 #2 #3 {J.\ Mol.\ Spectr.\ {\bf #1}, #3 (19#2)} % 
\def\jpa #1 #2 #3 {J.\ Phys.\ A {\bf #1}, #3 (19#2)}  % 
\def\jpsj #1 #2 #3 {J.\ Phys.\ Soc.\ Japan {\bf #1}, #3 (19#2)}
\def\lnc #1 #2 #3 {Lett.\ Nuov.\ Cim. {\bf #1}, #3 (19#2)}
\def\mj #1 #2 #3 {Math. Japon. {\bf #1}, #3 (19#2)} %  
\def\mpl #1 #2 #3 {Mod.\ Phys.\ Lett.\ A {\bf #1}, #3 (19#2)}
\def\nat #1 #2 #3 {Nature {\bf #1}, #3 (19#2)}
\def\nc #1 #2 #3 {Nuov.\ Cim.\ A{\bf #1}, #3 (19#2)}
\def\ncb #1 #2 #3 {Nuov.\ Cim.\ B{\bf #1}, #3 (19#2)}
\def\nima #1 #2 #3 {Nucl.\ Instr.\ and \ Meth.\ in\ Phys. \ Res. \ 
     A{\bf #1}, #3 (19#2)}
\def\nim #1 #2 #3 {Nucl.\ Instr.\ Meth.\ B{\bf #1}, #3 (19#2)}
\def\npb #1 #2 #3 {Nucl.\ Phys.\ B{\bf #1}, #3 (19#2)}
\def\pha #1 #2 #3 {Physica \ {\bf #1}, #3 (19#2)}
\def\pjm #1 #2 #3 {Pacific J. Math. \ {\bf #1}, #3 (19#2)} %  
\def\pla #1 #2 #3 {Phys.\ Lett.\ A {\bf #1}, #3 (19#2)}
\def\plb #1 #2 #3 {Phys.\ Lett.\ B {\bf #1}, #3 (19#2)}
\def\prep #1 #2 #3 {Phys.\ Rep. {\bf #1}, #3 (19#2)}  %
\def\pra #1 #2 #3 {Phys.\ Rev.\ A {\bf #1}, #3 (19#2)}  %
\def\prd #1 #2 #3 {Phys.\ Rev.\ D {\bf #1}, #3 (19#2)}
\def\prl #1 #2 #3 {Phys.\ Rev.\ Lett.\ {\bf #1}, #3 (19#2)}
\def\prs #1 #2 #3 {Proc.\ Roy.\ Soc.\ (Lon.) A {\bf #1}, #3 (19#2)}
\def\ptp #1 #2 #3 {Prog.\ Theor.\ Phys.\ {\bf #1}, #3 (19#2)}
\def\rmp #1 #2 #3 {Rev.\ Mod.\ Phys.\ {\bf #1}, #3 (19#2)} %
\def\ibid #1 #2 #3 {\it ibid., \rm {\bf #1}, #3 (19#2)}  %
\def\baps #1 #2 #3 {Bull.\ Am.\ Phys.\ Soc.\ {\bf #1}, #3 (19#2)} % 

\end{document}